\documentclass[american,english]{article}
\usepackage{mathpazo}
\usepackage{eulervm}
\usepackage[T1]{fontenc}
\usepackage[latin9]{inputenc}
\usepackage{xcolor}
\usepackage{calc}
\usepackage{enumitem}
\usepackage{amsmath}
\usepackage{graphicx}

\makeatletter

\providecommand{\tabularnewline}{\\}

\newlength{\lyxlabelwidth}      % auxiliary length 
	\newenvironment{elabeling}[2][]%
	{\settowidth{\lyxlabelwidth}{#2}
		\begin{description}[font=\normalfont,style=sameline,
			leftmargin=\lyxlabelwidth,#1]}
	{\end{description}}

\renewcommand\theparagraph    {%\thesubsubsection.%
\@arabic\c@paragraph}
\newcommand{\point}{\paragraph{\hskip-2.0ex.\hskip2.0ex}}

\AtBeginDocument{

}

\makeatother

\usepackage{babel}
\begin{document}
\title{\textbf{Intrinsic, extrinsic, and the constitutive }\textbf{\emph{a
priori}}\textbf{\textcolor{blue}{{} }}}
\author{László E. Szabó\\
\emph{\normalsize{}Department of Logic, Institute of Philosophy}\\
\emph{\normalsize{}Eötvös Loránd University Budapest}}
\date{~}
\maketitle
\begin{abstract}
On the basis of what I call physico-formalist philosophy of mathematics,
I will develop an amended account of the Kantian--Reichenbachian
conception of constitutive \emph{a priori}. It will be shown that
the features (attributes, qualities, properties) attributed to a real
object are not possessed by the object as a ``thing-in-itself'';
they require a physical theory by means of which these features are
constituted. It will be seen that the existence of such a physical
theory implies that a physical object can possess a property only
if other contingently existing physical objects exist; therefore,
the intrinsic--extrinsic distinction is flawed.
\end{abstract}

\section*{Introduction}

\point There is a long-standing debate in contemporary metaphysics
about the precise definition of \emph{intrinsic property}. The great
majority of the suggested definitions are some enhanced version of
Jaegwon Kim's (1982) definition, expressing the following simple idea: 
\begin{quote}
Intrinsic properties are the properties a particular object would
have even if no other contingently existing objects existed in the
world. (Allen~2018)
\end{quote}
Although Kim's definition has been widely accepted as basically adequate
criterion of intrinsicality, it has been challenged by subtle examples
and amended at several points. Beyond these improvements, the intrinsic--extrinsic
distinction is often discussed in a wider context of other closely
related metaphysical problems, such as categorical--dispositional,
pure--impure, relational--non-relational, interior--exterior distinctions;
the problem of identity, duplicate, persistence; or the problem of
Cambridge change. (E.g. Lewis 1983; Sider~1996; Humberstone~1996;
Vallentyne~1997; Langton and Lewis~1998; Francescotti~1999; Lewis~2001;
Marshall~2016; Marshall and Weatherson 2018). It is not my intention,
however, to review these debates, as the aim of this paper is to challenge
the concept of intrinsicality from a completely different point of
view and to show that the intrinsic--extrinsic distinction is flawed
on a more fundamental level. 

\point  Accordingly, my analysis will be restricted in two significant
senses:
\begin{enumerate}
\item The discussion will be restricted to the case of ordinary physical
properties of physical objects.
\item The whole analysis will remain within a radical physicalist ontological
doctrine: The world can be completely accounted for by assuming that
only physical entities exist. 
\end{enumerate}

\section*{Constitutive \emph{a priori}}

\point \label{point:Constitutive-a-priori}An object can possess
a property $X$ only if the term $X$ has meaning. This might sound
too radical. Probably, a less radical statement would be easily accepted:
The knowledge claim that an object possesses a property $X$ is possible
only if the term $X$ has meaning. But, just here is the philosophically
relevant point: this is not simply a matter of semantics. As Reichenbach
pointed out in his \emph{Relativity Theory and A Priori Knowledge}
(1920), the coordinative definitions, that is, the semantic \emph{conventions}
by which physical quantities or conceptions are defined in empirical
terms, play \emph{constitutive} role; they are constitutive \emph{a
priori}. ``They define the individual elements of reality and in
this sense \emph{constitute} the real object.'' (p.~53)

Reichenbach sharply distinguishes two different aspects of Kantian
\emph{a priori}: 
\begin{quotation}
\noindent Kant's concept of a priori has two different meanings. First,
it means ``necessarily true'' or ``true for all times,'' and secondly,
``constituting the concept of object.'' 

The second meaning must be clarified. According to Kant, the object
of knowledge, the thing of appearance, is not immediately given. Perceptions
do not give the object, only the material of which it is constructed.
Such constructions are achieved by an act of judgment. The judgment
is the synthesis constructing the object from the manifold of the
perception. For this purpose it orders the perceptions according to
a certain schema; depending on the choice of the schema, either an
object or a certain type of relation will result. Intuition is the
form in which perceptions present the material---thus performing
another synthesis. But the conceptual schema, the category, creates
the object; the object of science is therefore not a ``thing-in-itself''
but a reference structure based on intuition and constituted by categories.
(pp.~48--49) 
\end{quotation}
\point Let me give an example. Electrodynamics begins with the operational
definitions of the basic electrodynamic quantities. For example, electric
field strength is defined as  the force felt by the unit test charge
in electromagnetic field, when the test charge is at rest.\footnote{For the sake of simplicity I use this naive definition. For more precise
operational definitions of electrodynamic quantities, see Gömöri and
Szabó~2013.} This operational definition, therefore the notion of physical quantity
called electric field strength is \emph{a priori} in the sense that
it is prior to our empirical knowledge about the electromagnetic field.
There are no empirical facts which would determine the \emph{convention}
by which this physical quantity is defined. In other words, the electromagnetic
field, as a ``thing-in-itself'' does not determine that such a physical
quantity is necessarily introduced for its characterization, as one
of its fundamental feature. (In fact, the so called ``displacement
vector'', a different physical quantity having a completely different
operational definition, is an equally good alternative for the description
of electromagnetic field.) 

Once electric field strength is introduced, we can observe empirical
facts about it. We can measure, for example, the electric field strength
around a point charge being at rest, and, by inductive generalization,
we can ascertain the Coulomb Law: $E=\frac{kQ}{r^{2}}$. To be sure,
the law of physics we ascertained is \emph{a posteriori}; it can be
known only from empirical observations. But the \emph{features of
the physical reality} that the physical law talks about is \emph{a
priori} constructed by us. 

\point \label{point:argument} All this means that the features (attributes,
qualities, properties) attributed to a real object are not possessed
by the object as a ``thing-in-itself''; they require the existence
of something else, an epistemic agent constituting them. This recognition
outlines an argument against intrinsic--extrinsic distinction. 

\point  However, the Kantian--Reichenbachian account raises several
problems. It is not clear at all what kind of ontological entity is
involved by the required epistemic agency. Is it a contingently existing
flesh and blood physical being? Or is it an abstract entity, as sometimes
the Kantian ``transcendental subject'' is interpreted? (Carr~1999,
p.~53) To what extend is a semantic convention free? Are there ``coordinating
principles'', ``prescriptions for the conceptual side of the coordination'',
as Reichenbach (1920, p.~54) presupposes? And what is the origin
of such principles? Logic and mathematics? Are logical and mathematical
truths \emph{necessary }truths? If so, how to avoid then necessarily
true synthetic \emph{a priori} statements about the physical world?
Finally, how to incorporate semantic and epistemological holism, the
fact that ``our statements about the external world face the tribunal
of sense experience not individually but only as a corporate body''
(Quine~1951)?

In the next sections, I will attempt to formulate the argument outlined
in point~\ref{point:argument}, in such a way that the above mentioned
problems will be resolved or avoided. To achieve this aim, however,
I need to redraw a larger picture.

\section*{Physico-formalist account of logic and mathematics}

\point  I begin with a physicalist account of logic and mathematics.
How can the logical and mathematical facts be accommodated in a purely
physical ontology? Physicalism denies the existence of mental and
abstract entities; consequently, there is no room left for any kind
of platonism or mentalism in the philosophy of logic and mathematics.
Therefore, two possibilities are left: a Millian-style physical realist
approach, and formalism. We can disregard from the Millian realism,
in which mathematics itself becomes a physical theory, in the sense
of physical theory as described in the next section. So, formalism
-- more precisely, what I will call physco-formalism -- seems to
be the only account for logic and mathematics that can be compatible
with physicalism. Therefore, the formalist approach is our starting
point.

First we need to clarify: What are the logical/mathematical facts?
The formalist thesis can be summarized in one famous sentence of Hilbert\textbf{:}
``Mathematics is a game played according to certain simple rules with
meaningless marks.'' (Bell~1951, 38) Accordingly, in the formalist
account, a mathematical statement/fact/truth is like ``$\Sigma\vdash A$''.
According to the formalist view, neither $A$ nor the elements of
$\varSigma$ are statements, which could be true or false. They are
just meaningless strings, formulas of the formal system in question.
The entailment $\vdash$ relation has nothing to do with ``truth
preserving if-then-type reasoning''; it simply stands for the fact
that there is a finite sequence of meaningless strings of symbols,
fitting into some structural patterns called ``derivation rules''.
As a visual illustration, Fig.~\ref{fig:Csoportelm=0000E9let} shows
a short example: the first order axiomatic formulation of group theory
and the proof that $p(e,p(e,e))=e$ is a theorem. As this simple illustration
shows, such an ``evident truth'' that ``the unit element three
times multiplied by itself equals the unit element'' precisely means
the existence of a sequence of formulas, (1)--(9) in Fig.~\ref{fig:Csoportelm=0000E9let},
constituting a proof. 
\begin{figure}
\noindent \begin{centering}
\textsf{\textsc{\footnotesize{}}}%
\noindent{\fboxsep 10pt\fcolorbox{black}{white}{\begin{minipage}[t]{1\columnwidth - 2\fboxsep - 2\fboxrule}%

\subsubsection*{\textsf{\textsc{\footnotesize{}Alphabet}}}

\textsf{\textsc{\footnotesize{}}}%
\begin{tabular}{lll}
\textsf{\textsc{\footnotesize{}variables}} & \textsf{\textsc{\footnotesize{}$x,y,z,\ldots$}} & \tabularnewline
\textsf{\textsc{\footnotesize{}individual constants~~~~}} & \textsf{\textsc{\footnotesize{}$e$}} & \textsf{\textsc{\footnotesize{}(identity)}}\tabularnewline
\textsf{\textsc{\footnotesize{}function symbols}} & \textsf{\textsc{\footnotesize{}$i,p$}} & \textsf{\textsc{\footnotesize{}(inverse, product)}}\tabularnewline
\textsf{\textsc{\footnotesize{}predicate symbol}} & \textsf{\textsc{\footnotesize{}$=$}} & \tabularnewline
\textsf{\textsc{\footnotesize{}logical symbols}} & \textsf{\textsc{\footnotesize{}$\forall,\neg\rightarrow$}} & \tabularnewline
\textsc{\footnotesize{}others} & {\footnotesize{}$(,),\,,$} & \tabularnewline
\end{tabular}{\footnotesize\par}

\subsubsection*{\textsf{\textsc{\footnotesize{}Derivation rules}}}

\textsf{\textsc{\footnotesize{}}}%
\begin{tabular}{lll}
\textsf{\textsc{\footnotesize{}(MP)}} & \textsf{\textsc{\footnotesize{}$\left\{ \phi,\left(\phi\rightarrow\psi\right)\right\} $
$\vdash$ $\psi$}} & \textsf{\textsc{\footnotesize{}(modus ponens)}}\tabularnewline
\textsf{\textsc{\footnotesize{}(G)}} & \textsf{\textsc{\footnotesize{}$\phi$ $\vdash$ $\forall x\phi$ }} & \textsf{\textsc{\footnotesize{}(generalization)}}\tabularnewline
\end{tabular}{\footnotesize\par}

\subsubsection*{\textsf{\textsc{\footnotesize{}Axioms }}}

\textsf{\textsc{\footnotesize{}}}%
\begin{tabular}{ll}
\textsf{\textsc{\footnotesize{}(PC1)}} & \textsf{\textsc{\footnotesize{}$\phi\rightarrow\left(\psi\rightarrow\phi\right)$}}\tabularnewline
\textsf{\textsc{\footnotesize{}(PC2)}} & \textsf{\textsc{\footnotesize{}$\left(\phi\rightarrow\left(\psi\rightarrow\chi\right)\right)\rightarrow\left(\phi\rightarrow\psi\right)\rightarrow\left(\phi\rightarrow\chi\right)$}}\tabularnewline
\textsf{\textsc{\footnotesize{}(PC3)}} & \textsf{\textsc{\footnotesize{}$\left(\neg\phi\rightarrow\neg\psi\right)\rightarrow\left(\psi\rightarrow\phi\right)$}}\tabularnewline
\textsf{\textsc{\footnotesize{}(PC4)}} & \textsf{\textsc{\footnotesize{}$\forall x\left(\phi\rightarrow\psi\right)\rightarrow\left(\phi\rightarrow\forall x\psi\right)$~~(given
that $x$ is not free in $\phi$)}}\tabularnewline
\textsf{\textsc{\footnotesize{}(PC5)}} & \textsf{\textsc{\footnotesize{}$\forall x\phi\rightarrow\phi$~~(given
that $x$ is not free in $\phi$)}}\tabularnewline
\textsf{\textsc{\footnotesize{}(PC6)}} & \textsf{\textsc{\footnotesize{}$\forall x\phi(x)\rightarrow\phi(t)$~~(if
$t$ is a term which is free for $x$ in $\phi(x)$)}}\tabularnewline
\textsf{\textsc{\footnotesize{}(E1)}} & \textsf{\textsc{\footnotesize{}$x=x$}}\tabularnewline
\textsf{\textsc{\footnotesize{}(E2)}} & \textsf{\textsc{\footnotesize{}$t=s\rightarrow f^{n}\left(u_{1},u_{2},\ldots,t,\ldots u_{n}\right)=f^{n}\left(u_{1},u_{2},\ldots,s,\ldots u_{n}\right)$}}\tabularnewline
\textsf{\textsc{\footnotesize{}(E3)}} & \textsf{\textsc{\footnotesize{}$t=s\rightarrow\left(\phi\left(u_{1},u_{2},\ldots,t,\ldots u_{n}\right)\rightarrow\phi\left(u_{1},u_{2},\ldots,s,\ldots u_{n}\right)\right)$}}\tabularnewline
\textsf{\textsc{\footnotesize{}(G1)}} & \textsf{\textsc{\footnotesize{}$p(p(x,y),z)=p(x,p(y,z))$}}\tabularnewline
\textsf{\textsc{\footnotesize{}(G2)}} & \textsf{\textsc{\footnotesize{}$p(e,x)=x$}}\tabularnewline
\textsf{\textsc{\footnotesize{}(G3)}} & \textsf{\textsc{\footnotesize{}$p(i(x),x)=e$}}\tabularnewline
\end{tabular}{\footnotesize\par}

\subsubsection*{\textsf{\textsc{\footnotesize{}Theorem: ~~~ $p(e,p(e,e))=e$}}}

\textsf{\textsc{\footnotesize{}Proof:\medskip{}
}}{\footnotesize\par}

\textsf{\textsc{\footnotesize{}}}%
\begin{tabular}{lll}
\textsf{\textsc{\footnotesize{}(1)}} & \textsf{\textsc{\footnotesize{}$p(e,x)=x$}} & \textsf{\textsc{\footnotesize{}(G2)}}\tabularnewline
\textsf{\textsc{\footnotesize{}(2)}} & \textsf{\textsc{\footnotesize{}$\forall xp(e,x)=x$}} & \textsf{\textsc{\footnotesize{}(G)}}\tabularnewline
\textsf{\textsc{\footnotesize{}(3)}} & \textsf{\textsc{\footnotesize{}$\forall xp(e,x)=x\rightarrow p(e,e)=e$}} & \textsf{\textsc{\footnotesize{}(PC6)}}\tabularnewline
\textsf{\textsc{\footnotesize{}(4)}} & \textsf{\textsc{\footnotesize{}$p(e,e)=e$}} & \textsf{\textsc{\footnotesize{}(2), (3), (MP)}}\tabularnewline
\textsf{\textsc{\footnotesize{}(5)}} & \textsf{\textsc{\footnotesize{}$\forall xp(e,x)=x\rightarrow p(e,p(e,e))=p(e,e)$}} & \textsf{\textsc{\footnotesize{}(PC6)}}\tabularnewline
\textsf{\textsc{\footnotesize{}(6)}} & \textsf{\textsc{\footnotesize{}$p(e,p(e,e))=p(e,e)$}} & \textsf{\textsc{\footnotesize{}(2), (5), (MP)}}\tabularnewline
\textsf{\textsc{\footnotesize{}(7)}} & \textsf{\textsc{\footnotesize{}$p(e,e)=e\rightarrow p(e,p(e,e))=p(e,e)\rightarrow p(e,p(e,e))=e$~}} & \textsf{\textsc{\footnotesize{}(E3)}}\tabularnewline
\textsf{\textsc{\footnotesize{}(8)}} & \textsf{\textsc{\footnotesize{}$p(e,p(e,e))=p(e,e)\rightarrow p(e,p(e,e))=e$}} & \textsf{\textsc{\footnotesize{}(4), (7), (MP)}}\tabularnewline
\textsf{\textsc{\footnotesize{}(9)}} & \textsf{\textsc{\footnotesize{}$p(e,p(e,e))=e$}} & \textsf{\textsc{\footnotesize{}(6), (8), (MP)}}\tabularnewline
\end{tabular}{\footnotesize\par}%
\end{minipage}}}{\footnotesize\par}
\par\end{centering}
\caption{\emph{Group theory} \emph{and one of its theorems}\label{fig:Csoportelm=0000E9let}}
\end{figure}

\point  How can a formal system be accommodated in a purely physical
ontology? Where are the states of affairs located in the physical
world that make propositions like ``$\Sigma\vdash A$'' true or false?
For example, what are the facts of the physical world that make the
sequence of formulas (1)--(9) a proof of $p(e,p(e,e))=e$ in Fig~\ref{fig:Csoportelm=0000E9let}?
 Consider the first formula (1) in the proof. From ontological point
of view, what does it mean that it is nothing else but axiom (G2)?
It means that the formula $p(e,x)=x$ in row (1) is the same as the
formula $p(e,x)=x$ in row (G2). More exactly: it means the \emph{physical
fact} that the arrangement of black and white pixels on the screen
of my laptop on which I am writing this paper -- or the arrangement
of ink particles on the printout -- in row (1) and the corresponding
arrangement of black and white pixels in row (G2) are similar (congruent).
Similarly, what does it mean that formula $\forall xp(e,x)=x$ in
(2) is obtained from $p(e,x)=x$ by the derivation rule (G)? It means
that replacing the pixel-configuration $\phi$ in $\phi\vdash\forall x\phi$
, with copy/paste, by $p(e,x)=x$ in (1), the resulted pixel-configuration
is $p(e,x)=x\vdash\forall xp(e,x)=x$, such that the left hand side
is similar to the pixel-configuration $p(e,x)=x$ in (1) and the right
hand side is similar to $\forall xp(e,x)=x$ in (2). And so on and
so forth. 

Of course, a formal system can be thought of in different physical
forms; written with ink on paper, in the form of brain states and
brain processes, in the form of electronic states and electronic processes
in a computer, etc. What is important is the following observation:
\begin{quote}
\textbf{\emph{The physico-formalist thesis}}\emph{ ~~~The logical
and mathematical facts, since being formal facts, are nothing but
physical facts of physically existing formal systems consisting of
signs and derivation patterns embodied in concrete physical objects,
concrete physical configurations, and concrete physical processes.}
(Szabó~2003; 2012; 2017). 
\end{quote}
The physico-formalist account remains completely within the physicalist
ontology: there is no need for hypostatized entities -- abstract,
conceptual, or mental formal systems -- which exist over and above
the physically existing formal systems embodied in concrete physical
objects, concrete physical configurations, and concrete physical processes.

\point  ``Abstract'' or ``mathematical'' formal system is often
conceived as an entity obtained by \emph{abstraction, }by isolating
the common features in a number of particular physically existing
formal systems (e.g. Curry~1951, 30). To maintain the physico-formalist
thesis, it is worthwhile to illustrate that such an abstraction does
not lead out of the realm of the physically existing formal systems.
\begin{figure}
\noindent \begin{centering}
\includegraphics[width=0.6\columnwidth]{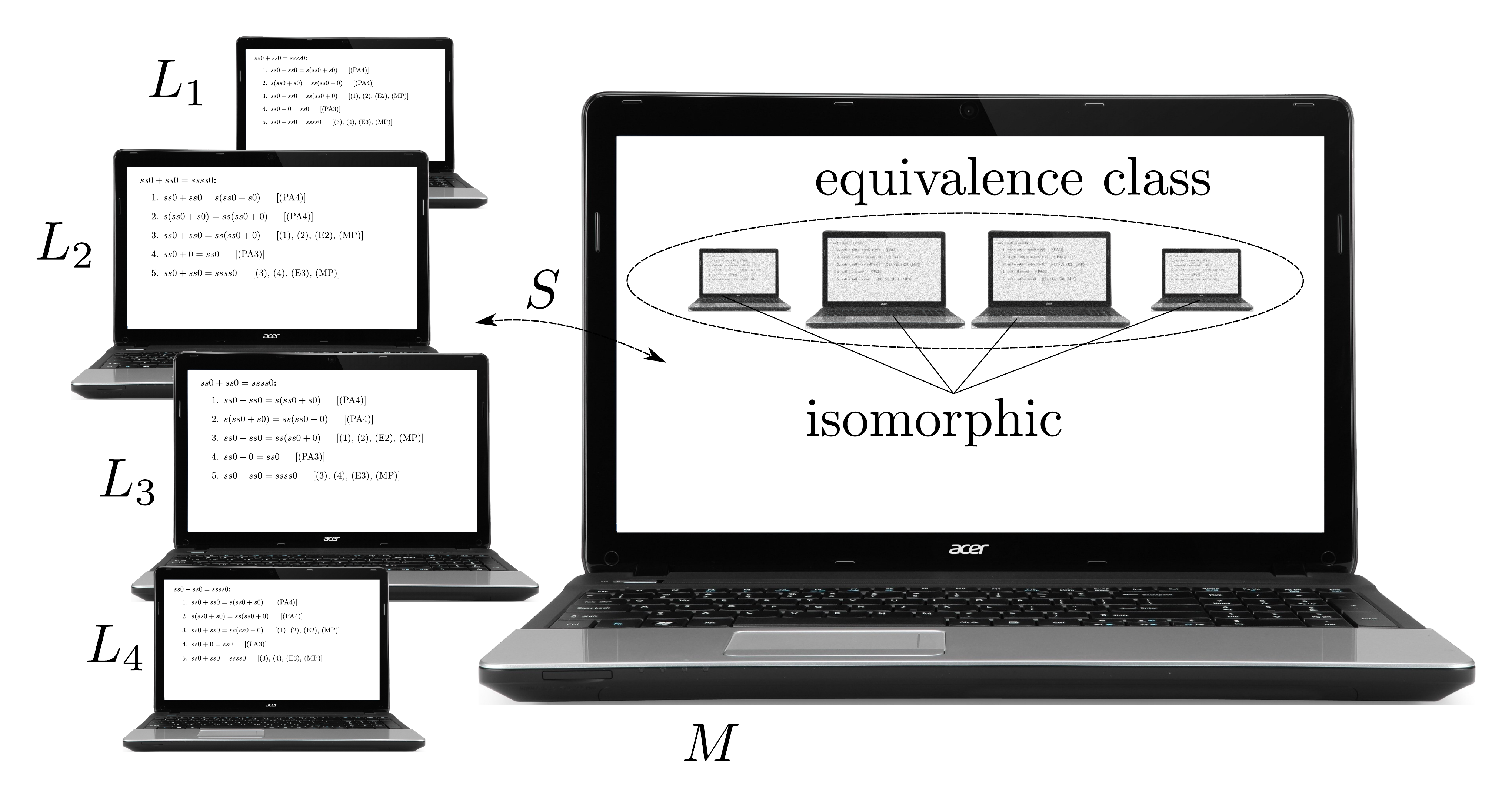}
\par\end{centering}
\caption{\emph{In order to isolate the common essential features of different
physically existing formal systems $L_{1},L_{2,},L_{3},\ldots L_{n}$
we must have a ``meta-theory'' of $L_{1},L_{2,},L_{3},\ldots L_{n}$,
using another physically existing formal system $M$}. \label{fig:meta-theory}}

\end{figure}

Consider a number of different physically existing formal systems
$L_{1},L_{2},\ldots L_{n}$, embodied in different particular physical
systems (Fig.~\ref{fig:meta-theory}). To abstract from some peculiar
properties of physical objects $L_{1},L_{2},\ldots L_{n}$, and to
isolate the common essential features of them, first of all requires
knowledge of the properties of the physical objects in question. That
is, we must have a physical theory $(M,S)$ -- in the sense of the
definition of physical theory in the next section -- describing $L_{1},L_{2,},L_{3},\ldots L_{n}$,
using \emph{another physically existing} formal system $M$. Abstraction
is a relationship, formulated in $(M,S)$, between two, a more detailed
and a less detailed, representations of $L_{1},L_{2,},L_{3},\ldots L_{n}$.
Only in a suitable formal system $M$ it is meaningful to talk about
similarity or isomorphism between the structures describing $L_{1},L_{2},\ldots L_{n}$,
and, for example, about the equivalence class of these structures,
which could be regarded as the product of the abstraction process.
But, all these are contained in $M$, a formal system existing in
the physical world. Thus, abstraction does not lead out of the physical
realm. It does not produce ``abstract formal systems'' over and
above the physically existing ones.\emph{ }One physically embodied
formal system can be applied in the description of another physically
embodied formal system, that's all.

\point Thus, any statement about a formal system -- including a
statement like ``$\Sigma\vdash A$'' -- is a statement of a physical
fact; consequently, it has exactly the same epistemological status
as any other statements about the physical world. This has far-reaching
consequences, of course: Logical and mathematical truths express objective
(mind independent) facts of a particular part of the physical world,
namely, the facts of the formal systems themselves. As such, they
are synthetic, \emph{a posteriori}, not necessary, and not certain;
they are fallible. But they have contingent factual content, as any
similar scientific assertion, so they ``can be true and useful and
surprising'' (Ayer~1952, 72). The logical and mathematical facts
can be discovered, like any other facts of nature, just like a fact
about a plastic molecule, or other artifact.

The fact that the formal systems usually are simple physical systems
of relatively stable behavior, like a clockwork, and that the knowledge
of logical and mathematical truths does not require observations of
the physical world external to the formal systems explains the universal\emph{
illusion} that logical and mathematical truths are necessary, certain
and \emph{a priori}. 

\section*{Physical theory}

\point \label{Point:Carnap}Following Carnap, a physical theory,
providing a description of a certain part of physical reality, $U$,
can be considered as a partially interpreted formal system, $(L,S)$.
The formal system $L$ is given by the language and the derivation
patterns, and the axioms. Traditionally, the axioms are divided into
the logical axioms (ideally, the first-order predicate calculus with
identity), the axioms of some mathematical theories, and, of course,
some physical axioms.

How can physical theory be accommodated in a purely physical ontology?
The $L$-part is already solved by the physico-formalist interpretation
of formal system and logical/mathematical fact. But, how can the physicalist
account for \emph{meaning} and \emph{truth}? Again, first we need
to clarify what it is that has to be accounted for; we need a definition
of semantic relationship between formulas of a formal system and states
of affairs in the physical world.

\point \label{point:semantics}The definition will be based on the
intuition we can learn from Gödel's construction of representation
of the meta-arithmetic facts in Peano arithmetic, in the preparation
of the first incompleteness theorem (e.g. Crossley~et~al.~1990,
62). In his construction, Gödel clearly defines, when we are entitled
to say that a formula of a formal system (in his case, a formula of
Peano arithmetic) represents/means/refers to a fact of the world outside
of the formal system (in his case, a meta-arithmetic fact). Mutatis
mutandis, we will repeat the same definition.

One is entitled to say that a formula $A$ represents or means a state
of affairs $a$ in $U$, if the following two conditions are met:
\begin{elabeling}{00.00.0000}
\item [{(A)}] There exist a family $\left\{ A_{\lambda}\right\} _{\lambda}$
of formulas in $L$ and a family $\left\{ a_{\lambda}\right\} _{\lambda}$
of states of affairs in $U$, such that $A=A_{\lambda_{0}}$ and $a=a_{\lambda_{0}}$
for some $\lambda_{0}$.
\item [{(B)}] For all $\lambda$,
\begin{eqnarray*}
\mbox{if }a_{\lambda}\mbox{ is the case in \ensuremath{U}} & \mbox{ then } & \Sigma\vdash A_{\lambda}\\
\mbox{if }a_{\lambda}\mbox{ is not the case in \ensuremath{U}} & \mbox{ then } & \Sigma\vdash\neg A_{\lambda}
\end{eqnarray*}
\end{elabeling}

\point \label{point: A-few-important -remarks}A few important remarks
are in order. 
\begin{elabeling}{00.00.0000}
\item [{(a)}] As we have seen, to be a meaning-carrier is not simply a
matter of convention or definition or declaration. Semantics \emph{is
not an arbitrary assignment} of states of affairs of the world to
linguistic elements of the theory.
\item [{(b)}] It is pointless to talk about the meaning of an \emph{isolated}
formula of the theory. (Semantic holism) It is not only because of
condition (A), but also because in condition (B) a big part of the
axiomatic system can be involved.
\item [{(c)}] It must be recognized that condition (B) is exactly the same
as the necessary and sufficient condition for the theory $(L,S)$
to be \emph{true}. That is, the two conceptions \emph{meaning and
truth are completely intertwined}. 
\item [{(d)}] The semantics, in the above holistic sense, is a part and
parcel of physical theory. In case of empirical failure of a physical
theory, semantics is one of the possible object of revision. In other
words, semantics is as much hypothetical as any other part of the
theory. 
\end{elabeling}
\point \label{point:commoncause}It must be clear that $a_{\lambda}$
-- as a symbol in the meta-language we use to describe the semantic
relationship -- stands for a state of affairs, a configuration of
\emph{the physical world}. Also, according to the physico-formalist
approach ``$\Sigma\vdash A_{\lambda}$'' and ``$\Sigma\vdash\neg A_{\lambda}$$"$,
respectively, are states of affairs in \emph{the physical world},
facts of the physically embodied  formal system $L$. Thus, what we
observe in condition (B) is a kind of regularity or \emph{correlation
between physical facts} of two parts of the physical world, $L$ and
$U$. Combining this with the thesis of \emph{the principle of common
cause}\footnote{I mean the Reichenbachian thesis that no correlation without causation;
every correlation is either due to a direct causal effect, or is brought
about by a third factor, a so-called common cause (e.g. Reichenbach~1956;
Hofer-Szabó~et al.~2013).}, one must conclude that both semantic relationship and the truth
of the physical theory (consequently, our \emph{knowledge}) must be
brought about by the underlying causal processes of the physical world,
going on in the common causal past of the two parts of the world $L$
and $U$. This underlying process is what we normally call \emph{learning
through experience}. That is, \emph{no genuine knowledge of the physical
world is possible without experience}. By the same token, \emph{no
semantically meaningful talk about the physical world is possible
without experience}. There is no \emph{a priori} meaning and there
is no \emph{a priori} truth.

\point Let us clarify then the role of logic and mathematics in our
knowledge of the physical world. One might ask: if mathematics is
only about the formal systems without meaning, how is it, then, possible
that mathematical structures prove themselves to be so expressive
in the physical applications? As Richard Feynman put it: ``I find
it quite amazing that it is possible to predict what will happen by
mathematics, which is simply following rules which really have nothing
to do with the original thing.'' (\foreignlanguage{american}{Feynman~1967,
171) }

We need to separate the sociological/practical aspect of this issue
from the epistemological one. Let me start with the sociological aspect.
From sociological point of view, by ``mathematics'' we mean that
part of the axioms of $L$ (see point~\ref{Point:Carnap}) which
is traditionally considered as non-physical axioms. It must be clear
however that it is not ``mathematics'' \emph{alone} by which the
physicist can predict what will happen, but the physical axioms together
with the logical and mathematical axioms. The physicist, keeping,
as long as possible, the logical and mathematical axioms fixed --
for good sociological/practical reason --, \emph{tunes} the physical
axioms such that the theorems \emph{derivable} from the unified system
of logical, mathematical, and physical axioms (together with the deduction
rules) + the semantics, as a whole, be compatible with the empirical
facts. Consequently, the employed logical and mathematical structures
in themselves need not, and do not, reflect anything about the real
world in order to be useful. 

Let me recall an analogy I gave in my (2017). You can experience a
analogous situation when you change the mouse driver on your computer
(or just change the mouse settings): first you feel that the pointer
movements (``derived theorems'') generated by the new driver (``mathematics'')
according to your previously habituated hand movements (``physical
axioms'') do not faithfully reflect the arrangement of your screen
content (experienced world). Then, keeping the driver (and driver
settings) fixed, you \emph{tune} your hand movements -- through typical
``trial and error'' learning -- such that the generated pointer
movements fit to the arrangement of your screen content. 

Thus, there is no miraculous ``preadaption'' or ``pre-established
harmony'' involved just because certain aspects of empirical reality
``fit themselves into the forms provided by mathematics''. This
is simply a result of the physicist's continuous choice from the store
shelves of mathematics, and the continuous tuning the physical axioms. 

From epistemological point of view the situation is even more simple.
In a physical theory $(L,S)$, the formal system $L$ is a single
undivided whole. As a formal system, $L$ can be the object of logical/mathematical
investigation and knowledge. In itself, however, it has nothing to
do with the physical world described by the \emph{physical theory}
$(L,S)$. In case of empirical failure of the physical theory $(L,S)$,
any element of $L$ as well as the semantics can be the object of
revision. Through this continuous ``trial and error'' process, the
physicist tunes the physical theory $(L,S)$, as a whole, be compatible
with the experienced world; satisfy condition (B) in point~\ref{point:semantics}.

\section*{Constitutive \emph{a priori }reconsidered}

\point Consider the following example. Given that $L$ is consistent,
one can easily see that the following statements cannot hold true
at the same time:
\begin{elabeling}{00.00.0000}
\item [{(i)}] $A$ refers to $a$
\item [{(ii)}] $L\vdash A$
\item [{(iii)}] $a$ is not the case in $U$
\end{elabeling}
since (i) and (iii), according to condition (B) in point~\ref{point:semantics},
would imply $L\vdash\neg A$, in contradiction with (ii). Therefore,
observing that $a$ is not the case we are not entitled to say that
we observe that ``$\neg A$''. Simply because if $a$ is not the
case, then condition (B) fails, and the whole semantics is lost. Therefore
$\neg A$ does not carry meaning at all. That is to say, we are not
able to attribute a feature, whatsoever, to the physical reality in
the situation when $a$ is not the case. 

\point  \label{point:dinamikus}Let us denote this unexpressed, unarticulated
state of affairs by $a^{*}$. Once a modified (or completely new)
theory, $(L',S')$, is constructed with a new family of state of affairs
$\left\{ a'_{\lambda'}\right\} _{\lambda'}$ and a new family of formulas
$\left\{ A'_{\lambda'}\right\} _{\lambda'}$, such that $a^{*}=a'_{\lambda'_{*}}$
and condition (B) in point~\ref{point:semantics} is satisfied, the
corresponding $A'_{\lambda'_{*}}$ will be attributed to $a^{*}$,
as true feature of reality in state $a^{*}$.

\point  This example not only confirms what was said in point~\ref{point: A-few-important -remarks}~(d),
but also sheds light on the constitutive role of semantics, similar
to the constitutive role of Reichenbach's coordinative definitions
we described in point~\ref{point:Constitutive-a-priori}. There are
however significant differences:
\begin{elabeling}{00.00.0000}
\item [{(a)}] The whole account I developed remains within the clear and
minimal ontological framework of physicalism.
\item [{(b)}] All entities involved are contingently existing physical
objects.
\item [{(c)}] Whatever is the concrete physical process (point~\ref{point:commoncause})
producing the correlation required in condition (B) of point~\ref{point:semantics},
that is, producing meaning and truth, it is a contingently existing
part of the physical reality.
\item [{(d)}] Therefore, the whole ``epistemic agency'' involved is embodied
in the physical world.
\item [{(e)}] The formal system in $(L,S)$ plays, indeed, a similar role
as Reichenbach's ``constitutive principle''; except that
\begin{itemize}
\item $L$ has a clear ontological status; it is a part of the physical
world,
\item and all facts of $L$ are contingent facts of the physical world,
therefore they do not generate necessarily true synthetic \emph{a
priori} statements about the physical world, for sure.
\end{itemize}
\item [{(f)}] There is no ``conceptual side of the coordination''. The
constitutive role of formal systems by no means entitles us to say
that there is a hypostatized \emph{a priori} conceptual scheme in
terms of which we grasp the experienced reality, and that this conceptual
scheme generates analytic truths. For, what there is is anything but
 conceptual: we only have the physically existing formal systems which
have no meaning. Once an otherwise meaningless formula of a formal
system is provided with meaning, in the sense of point~\ref{point:semantics},
it becomes true or false in a non-analytic sense.
\item [{(g)}] Certainly, semantics plays constitutive role; not in the
form of isolated operational definitions of physical concepts, but
in holistic sense. Moreover, semantics is completely intertwined with
the truth of the theory; also, not in the form of the truth of isolated
predictions, but in holistic sense of (B) in point~\ref{point:semantics}. 
\item [{(h)}] In fact, therefore, what plays the constitutive role is the
whole theory $(L,S)$, though, the whole theory is certainly not \emph{a
priori}. As we have seen, however, aprioricity is not required for
playing a constitutive role. That is why there is no tension, whatsoever,
between the fact that the applied logical and mathematical structures
as well as the constructed semantics can \emph{change} (see point~\ref{point:dinamikus})
when a theory is superseded by another one, on the one hand, and their
constitutive role in furnishing reality, on the other (cf. Friedman~2001;
MacArthur~2008; Ivanova~2011). 
\item [{(i)}] Finally, it is worthwhile to mention that the constitutive
role of the theory $(L,S)$ in furnishing ``the »continuum« of reality''
(Reichenbach~1920, p.~50) with objects, properties, and relations,
in itself, by no means implies the denial of realism; the belief that
the existence of the objects, properties, and relations posited by
the theory is a real feature of reality. What is true about scientific
realism is also true about metaphysical realism -- if the distinction
is meaningful at all. For a ``metaphysical'' account must have the
structure of $(L,S)$ in order to be meaningful and true, just in
the sense we described in point~\ref{point:semantics}. 
\end{elabeling}
\point Thus, the corrected formulation of the argument against intrinsic--extrinsic
distinction, outlined in point~\ref{point:argument}, is the following.
The features (attributes, qualities, properties) attributed to a real
object are not possessed by the object as a ``thing-in-itself'';
they require the existence of something else: a physical theory $(L,S)$
by means of which these features are constituted. The existence of
$(L,S)$ however implies the existence of a physically embodied formal
system $L$ and a real causal process in the physical world producing
the correlation required for both the semantics and the truth of the
theory. All this means that a physical object can possess a property
only if other contingently existing physical objects exist. 

\section*{Acknowledgments}

Funding was provided by Hungarian National Research, Development and
Innovation Office (Grant No. K115593). 

\section*{References}
\begin{elabeling}{x.xx.xx}
\item [{Allen,~Sophie~(2018):}] Properties, \emph{The Internet Encyclopedia
of Philosophy}, https://www.iep.utm.edu/properties.
\selectlanguage{american}%
\item [{Ayer,~Alfred~J.~(1952):}] \emph{Language,Truth and Logic. }New
York: Dover Publications.
\selectlanguage{english}%
\item [{Bell,~Eric~T.~(1951):}] \emph{Mathematics: Queen and Servant
of Science.} New York: McGraw-Hill Book Company.
\item [{Crossley,~J.N.,~C.J.~Ash,~Stillwell,~J.C.,~Williams,~N.H.,~and~Brickhill,~C.J.~(1990):}] \emph{What
is mathematical logic? }New York: Dover Publications.
\item [{Carr,~David~(1999):}] \emph{The Paradox of Subjectivity. The
Self in the Transcendental Tradition}. Oxford: Oxford University Press.
\item [{Curry,~Haskell~B.~1951.}] \emph{Outlines of a Formalist Philosophy
of Mathematics. }Amsterdam: North-Holland.
\selectlanguage{american}%
\item [{Feynman,~Richard.~(1967):}] \emph{The character of physical law.
}Cambridge: MIT Press.
\selectlanguage{english}%
\item [{Francescotti,~Robert~(1999):}] How to define intrinsic properties,
\emph{Noûs} \textbf{33}, 590--609. 
\item [{Friedman,~Michael~(2001):}] \emph{Dynamics of reason}. Stanford
(CA): CSLI Publications. 
\item [{Gömöri,~Márton,~and~Szabó,~László~E.~(2013):}] Operational
understanding of the covariance of classical electrodynamics, \emph{Physics
Essays }\textbf{26},  361--370.
\item [{Hofer-Szabó,~Gábor,~Miklós~Rédei,~and~László~E.~Szabó.~(2013):}] \emph{The
Principle of the Common Cause.} Cambridge: Cambridge University Press.
\item [{Humberstone,~Lloyd~(1996):}] Intrinsic/extrinsic, \emph{Synthese}
\textbf{108}, 205--267. 
\item [{Ivanova,~Milena~(2011):}] Friedman\textquoteright s Relativised
A Priori and Structural Realism: In Search of Compatibility, \emph{International
Studies in the Philosophy of Science} \textbf{25}, 23-37.
\item [{Langton,~Rae~and~Lewis,}] David~(1998): Defining `intrinsic\textquoteright ,
\emph{Philosophy and Phenomenological Research} \textbf{58}, 333--345. 
\item [{Lewis,~David~(1983):}] Extrinsic properties, \emph{Philosophical
Studies} \textbf{44}, 197--200. 
\item [{Lewis,~David~(2001):}] Redefining `Intrinsic', \emph{Philosophy
and Phenomenological Research} \textbf{63}, 381--398. 
\item [{Marshall,~Dan~(2016):}] An analysis of intrinsicality, \emph{Noûs}
\textbf{50}, 704--739. 
\item [{Marshall,~Dan~and~Weatherson,~Brian~(2018):}] Intrinsic vs.
Extrinsic Properties, \emph{The Stanford Encyclopedia of Philosophy}
(Spring 2018 Edition), Edward N. Zalta (ed.), https://plato.stanford.edu/archives/spr2018/entries/intrinsic-extrinsic.
\item [{McArthur,~Dan~(2008):}] Theory change, structural realism, and
the relativised \emph{a priori}, \emph{International Studies in the
Philosophy of Science} \textbf{22}, 5--20. 
\item [{Quine,~W.~V.~(1951):}] Two Dogmas of Empiricism, \emph{Philosophical
Review} \textbf{60}, 20-43.
\item [{Reichenbach,~Hans~(1920):}] \emph{The Theory of Relativity and
A Priori Knowledge}. Berkeley and Los Angeles: University of California
Press, 1965.
\item [{Reichenbach,~Hans.~(1956):}] \emph{The Direction of Time.} Berkeley:
University of California Press.
\item [{Sider,~Theodore~(1996):}] Intrinsic Properties, \emph{ Philosophical
Studies} \textbf{83}, 1--27. 
\item [{Szabó,~László~E.~(2003):}] Formal System as Physical Objects:
A Physicalist Account of Mathematical Truth, \emph{International Studies
in the Philosophy of Science}\textbf{ 17}, 117--125.
\item [{Szabó,~László~E.~(2012):}] Mathematical facts in a physicalist
ontology, \emph{Parallel Processing Letters}\textbf{ 22}, 1240009.
\item [{Szabó,~László~E.~(2017):}] Meaning, Truth, and Physics, In G.
Hofer-Szabó, L. Wro{\'n}ski (eds.), \emph{Making it Formally Explicit},
European Studies in Philosophy of Science 6, Berlin: Springer International
Publishing.
\item [{Vallentyne,~Peter~(1997):}] Intrinsic properties defined,\emph{
Philosophical Studies} \textbf{88}, 209--219.
\end{elabeling}

\end{document}